\documentclass[aps,prl,twocolumn,amsfonts,amsmath,amssymb,showpacs,final,10pt]{revtex4-1}
\usepackage[dvipsnames]{xcolor}
\usepackage[colorlinks=true,citecolor=Blue,linkcolor=Blue,urlcolor=Blue]{hyperref}
\usepackage{graphicx,manfnt,pifont,colonequals}
\usepackage{bbm}

\newcommand\noi{\noindent}

\newcommand\restr[2]{{\left.\kern-\nulldelimiterspace#1\vphantom{\big|}\right|_{#2}}}

\begin{document}
\title{Schur correlation functions from \texorpdfstring{$q$}{}-deformed Yang-Mills}
\author{Yufan Wang, Yiwen Pan} 
\affiliation{School of Physics, Sun Yat-Sen University, Guangzhou 510275, China}

\begin{abstract}
\noi We construct the wave functions in the $q$-deformed 2d Yang-Mills theory that compute torus correlation functions of affine currents in the VOA associated to a class of 4d $\mathcal{N} = 2$ SCFTs. These wave functions are then shown to reduce to the topological correlators of a set of Coulomb branch operators in the $T[SU(N)]$ theory, from which those correlators in the 3d mirror dual of the 4d $T_N$ theories can be computed.
\end{abstract}

\maketitle

\section{Introduction}

Four dimensional $\mathcal{N} = 2$ SCFTs are found to contain a privileged class of operators, referred to as the ``Schur operators'', endowed with a vertex operator algebra (VOA) structure \cite{Beem:2013sza}. The $\mathcal{R}$-symmetry of the parent theory guarantees the presence of a Virasoro subalgebra, while any flavor symmetry $\mathfrak{g}$ enhances to an affine algebra $\widehat{\mathfrak{g}}_k$, whose generators descend from the flavor moment maps in 4d. The full VOAs of a variaty of 4d SCFTs have been identified and analyzed in great depth, see for example \cite{Beem:2014rza,Lemos:2014lua,Song:2017oew,Xie:2019yds}. The underlying TQFT that captures the Schur index $\mathcal{I}$ of a 4d $\mathcal{N} = 2$ SCFT is identified with the $q$-deformed 2d Yang-Mills theory, which allows one to identify the index $\mathcal{I}$ as a correlator in the latter 2d theory \cite{Gadde:2011ik}. 

Recently, correlation functions of Schur operators on the torus are computed via supersymmetric localization \cite{Pan:2019bor,Dedushenko:2019yiw} and are further mapped to the deformation quanzation of the Higgs branch of vacua \cite{Beem:2016cbd,Dedushenko:2016jxl,Dedushenko:2019mzv,Pan:2019shz,Dedushenko:2019mnd,Panerai:2020boq} of the circle-reduced theory in 3d. The aim of this letter is to propose a parallel approach to computing some of these correlators, by constructing the wave functions corresponding to the affine current operators for a class of 4d SCFTs and show that they are determined by the correlators of critical current algebras. Moreover, we will relate the dimensionally-reduced wave functions for the $T_N$ theories with the Coulomb branch correlators in the 3d mirror dual \cite{Benini:2010uu,Dedushenko:2017avn}.

\section{Current correlation functions}


Let us begin with class $\mathcal{S}$ theories with only regular punctures. First we recall that each full puncture of such a theory $\mathcal{T}$ is associated to a factor of flavor symmetry $\mathfrak{g}$ with flavor central charge $k^\text{4d} = 2h^\vee$. Under the map from Schur operators in 4d to the vertex operators in the associated VOA $V \equiv \operatorname{VOA}[\mathcal{T}]$, the moment map operators of $\mathfrak{g}$ descend to the affine current generators of a current subalgebra $\widehat{\mathfrak{g}}_{k = -h^\vee}$ at the critical level.

The Schur index $\mathcal{I}(\mathbf{x},q)$, which is equivalent to the vacuum character $\operatorname{ch}_V(\mathbf{x},q)$ of $V \equiv \operatorname{VOA}[\mathcal{T}]$, can be written as a 2d $q$-deformed Yang-Mills correlators \cite{Beem:2014rza,Gadde:2011ik}
\begin{align}
  & \ \operatorname{ch}_V(\mathbf{x}, \mathbf{y}; q)
  \equiv \operatorname{tr}_{V}q^{L_0 - \frac{c}{24}} \mathbf{x}^\mathbf{h} \mathbf{y}^\mathbf{h'}\\
  = & \ q^{- \frac{c}{24}}\sum_\lambda C_\lambda(q)^{2 g - 2 + s} q^{(\lambda, \rho)} \operatorname{ch}_\lambda^\text{crit}(\mathbf{x}, q) \prod_{i = 1}^{s}\psi_\lambda^{(i)}(\mathbf{y}, q) \nonumber\ .
\end{align}
Here, the sum over $\lambda$ exhausts all integral dominant weights $\lambda$ of the finite Lie algebra $\mathfrak{g}$, and $\operatorname{ch}^\text{crit}_\lambda(\mathbf{x}, q)$ denotes the $\widehat{\mathfrak{g}}_{k = - h^\vee}$-character of a module whose highest weight subspace is given by the $\mathfrak{g}$-irreducible representation $\mathcal{R}_\lambda$ with highest weight $\lambda$ \cite{2007arXiv0706.1817A}. The appearance of the critical character is due to the simple relation $\operatorname{ch}_\lambda^\text{crit}(\mathbf{x}, q) =q^{ - (\lambda, \rho)}C_\lambda(q)\psi^\text{full}_\lambda(\mathbf{x}, q) $ with $C^{-1}_\lambda(q) \equiv \operatorname{PE}[\sum_{j = 1}\frac{q^{d_j}}{1-q}] \dim_q \mathcal{R}_\lambda$ and $\psi^\text{full}_\lambda \equiv \operatorname{PE}[\frac{1}{1-q}q\chi_\text{adj}^\mathfrak{g}]\chi_\lambda$ \cite{Lemos:2014lua}. The remaining $\psi_\lambda^{(i)}$ in the formula denotes the wave functions of other regular punctures. Our goal is to derive the torus correlation functions (in the unflavoring limit $\mathbf{x} \to 1$) of the currents $J^a$ associated to a full puncture.

Obviously $J^a$ have vanishing one-point functions. For the two-point functions, we recall Zhu's recursion relations \cite{10.2307/2152847} (see also section 3 there for definitions of various special functions), now applied to a module $M$ of the algebra $V$ and the generators $J^a$ of the vertex subalgebra $\widehat{\mathfrak{g}}_k$,
\begin{align}\label{2ptrecursion}
  \operatorname{tr}_M & J^a[z_1]J^b[z_2]q^{L_0} = \operatorname{tr}_M o(J^a)o(J^b)q^{L_0} \\
    & \ - \sum_{m \ge 0} (-1)^{m} P_{m + 1}(z_1 - z_2)\operatorname{tr}_M o(J^a[m]J^b) q^{L_0} \ .\nonumber
\end{align}
Here, for any state $a \in V$, the state-operator correspondence is given by $a[z] \equiv Y(e^{2\pi i z L_0}a, e^{2\pi i z}) =  e^{2\pi i \text{wt}(a) z}a(e^{2\pi i z})$ appropriate for insertions on a torus. The operator $a[m]$ denotes the square-bracket mode of $a$ given by
\begin{align}
  a[m] \equiv (2\pi i)^{- m - 1} \sum_{i \ge m} c(m, i, \operatorname{wt}(a))a_i \ ,
\end{align}
with $c(m, i, h)$ defined via $[\ln (1 + z)]^m ( 1 + z)^{\operatorname{wt}a - 1} = \sum_{ i \ge m} c(m, i, \operatorname{wt}a)z^i$. In the second term of (\ref{2ptrecursion}), only the summand with $m = 1$ contributes thanks to the commutation relations for $\widehat{\mathfrak{g}}_k$. Therefore
\begin{align}\label{twopointrecursion}
  & \ \operatorname{tr}_M J^a[z_1]J^b[z_2]q^{L_0} \\ 
  = & \ \operatorname{tr}_M o(J^a)o(J^b)q^{L_0} + \frac{k K^{ab}}{(2\pi i)^2} P_2(z_1 - z_2) \operatorname{tr}_M q^{L_0}  \ . \nonumber
\end{align}
Here $K$ denotes the Killing form. Group theory ensures that the first term on the right can be written as $K^{ab}\mathcal{J}(q)$. In particular, with a basis of Cartan generators $h^i$ for the finite part $\mathfrak{g}$ of $\widehat{\mathfrak{g}}_k$, we have
\begin{align}
  K^{jj}\mathcal{J}_M(q) = \left(x_{j} \frac{d}{dx_{j}}\right)^2\Bigg|_{x \to 1} \operatorname{tr}_M q^{L_0} \prod_{i = 1}^{r} x_{i}^{h^i} \ .
\end{align}

Therefore we conclude that the current two-point functions in any module $M$ can be obtained from the associated character by acting with a uniform differential operator, $\operatorname{tr}_M J^a[z_1]J^b[z_2]q^{L_0 - c/24} = \mathcal{D}^{ab}(z_1, z_2) \operatorname{ch}_M(q) $,
\begin{align}
   \mathcal{D}^{ab}(z_i) \equiv K^{ab}\left[\frac{1}{K^{ij}}\left(x_{j} \frac{d}{dx_{j}}\right)^2_{x\to 1} + k P_2(z_1 - z_2)\right] \ . \nonumber
\end{align}
The action of this differential operator on the Schur index leads easily to
\begin{align}
  & \ \operatorname{tr}_V J^a[z_1]J^b[z_2] q^{L_0} \\
  = & \ \sum_\lambda C_\lambda^{2g -2 + s} q^{(\lambda, \rho)} \operatorname{tr}_\lambda J^a[z_1]J^b[z_2]q^{L_0^\text{crit}} \prod_{i = 1}^{s} \psi^{(i)} \ . \nonumber
\end{align}
Said differently, the wave function $\Psi_\lambda^{ab}(z_1, z_2)$ in the $q$-deformed Yang-Mills theory corresponding to the two-point current insertions in $V$ is simply the (normalized) critical current two-point function
\begin{align}
  \Psi^{ab}_\lambda(z_1, z_2) = \langle J^a[z_1]J^b[z_2]\rangle^\text{crit}_\lambda \equiv \frac{\operatorname{tr}_\lambda J^a[z_1]J^b[z_2]q^{L_0^\text{crit}}}{\operatorname{tr}_\lambda q^{L_0^\text{crit}}} \ . \nonumber
\end{align}
In fact, one can do better by observing, for critical currents algebras, that
\begin{align}\label{Jlambda}
  \mathcal{J}_\lambda(q) = \left[\frac{|\lambda + \rho|^2 }{\dim \mathfrak{g}} + h^\vee E_2(\tau) \right] \operatorname{ch}_\lambda^\text{crit}(q)\ .
\end{align}
To prove this observation, it is convenient to rewrite the formal character $\operatorname{ch}_\lambda^\text{crit}$ evaluated at $2\pi i t\rho$ as
\begin{align}
  \operatorname{ch}_\lambda^\text{crit}(2\pi i t\rho)= f(q) \prod_{\alpha > 0} \frac{e^{\pi i t (\lambda + \rho, \alpha) } - e^{ - \pi i t (\lambda + \rho, \alpha) }}{\vartheta_1(t(\rho, \alpha|\tau))} \ .
\end{align}
Now we take $\partial_t^2|_{t = 0}$ with proper normalization by the Killing form. Using $\sum_{\alpha > 0}(\mu, \alpha)(\alpha, \nu) = h^\vee (\mu, \nu)$ and
\begin{align}
  E_2 = \frac{1}{12\pi^2} \frac{\vartheta_1'''(0|\tau)}{\vartheta_1'(0|\tau)} \ ,
\end{align}
one easily proves the statement. Explicitly, we have the wave function
\begin{align}
  \Psi^{ab}_\lambda(z_i) = -\frac{ h^\vee K^{ab}}{(2\pi i)^2}\wp_0(z_1 - z_2, \tau) + K^{ab}\frac{|\lambda + \rho|^2}{\dim \mathfrak{g}} \ .
\end{align}
Here we made use of the relation between $P_2$ and the Weierstrass $\wp$-function.

Next we turn to the three-point functions. To proceed, we recall Zhu's genreal recursion relations, proposition 4.3.4 of \cite{10.2307/2152847}. Again, substituting in $a = J^a, a^{(i)} = J^{a_i}$ we have
\begin{align}
  & \ \operatorname{tr}_M J^a[w] J^{a_1}[z_1]J^{a_2}[0]q^{L_0} \nonumber\\
  = & \ \operatorname{tr}_M (J^a[-1]J^{a_1})[z_1] J^{a_2}[0]q^{L_0}  + \frac{1}{2} \operatorname{tr}_M (J^a_0J^b)[z_1] J^c[0]q^{L_0}\nonumber\\
    & \ + \frac{i}{2\pi} \bigg[ (P_1(- z_1) - P_1(- w) ) \operatorname{tr}_M J^{a_1}[z_1](J^a_0J^{a_2})[0]q^{L_0} \nonumber\\
    & \qquad \qquad \qquad - P_1(z_1 - w) \operatorname{tr}_M (J^a_0J^{a_1})[z_1]J^{a_2}[0]q^{L_0} \bigg] \ .\nonumber
\end{align}
Let us first focus on Lie algebras $\mathfrak{g}$ with a vanishing cubic Casimir. Combining with $J^a_0J^b = i f^{ab}{_c} J^c$ and the Ward identity
\begin{align}
  & \ \operatorname{tr}_M (J^a[-1]J^{a_1})[z_1] J^{a_2}[0]q^{L_0} \nonumber
  \\
  = & \  - \frac{1}{2} f^{a a_1}{_b} \partial_{z_1} \operatorname{tr}_M J^b[z_1]J^{a_2}[z_2] q^{L_0} \ ,
\end{align}
we arrive at a simple recursion relation
\begin{align}
  \operatorname{tr} J^a[w] \prod_{i = 1}^2 J^{a_i}[z_i]q^{L_0}
  = \mathcal{D}^{a a_1}{_b} \operatorname{tr} J^b[z_1] J^{a_2}[z_2]q^{L_0} \nonumber
\end{align}
in terms of a uniform operator
\begin{align}
  \mathcal{D}^{a a_1}{_b} \equiv & \frac{1}{2} f^{a a_1}{_b}\bigg[
    - \partial_{z_1} + 1 \\
   & + \frac{1}{\pi i} (P_1(z_2 - z_1) - P_1(z_2 - w) + P_1(z_1 - w))
  \bigg]. \nonumber
\end{align}
We will also use the notation $D^{a_1a_2a_3}$ with a raised last index using the Killing form. Applying the operator $D^{a_1 a_2}{_b}$ to the two-point correlators of $V$, we therefore conclude that the wave function corresponding to three-point current insertions is simply the normalized three-point current correlator,
\begin{align}\label{3ptwavefunction}
  \Psi^{a_1 a_2 a_3}_\lambda(z_1, z_2, z_3) = \langle J^{a_1}[z_1]J^{a_2}[z_2]J^{a_3}[z_3]\rangle^\text{crit}_\lambda \ .
\end{align}

For $\mathfrak{g} = \mathfrak{su}(n)$ with $n \ge 3$, an additional term proportional to the cubic Casimir $d^{a_1 a_2 a_3}$ appears \cite{Dolan:2007eh}. This term is independent of $z_i$ and can be obtained by computing  $\operatorname{tr}_Vo(J^{(a_3})o(J^{a_2})o(J^{a_3)}) q^{L_0}$ with $a_i$ symmetrized. As shown in \cite{Dolan:2007eh}, the term can also be written as a uniform differential operator acting on the character. Therefore the above argument goes through and the result (\ref{3ptwavefunction}) holds. It will be important later to observe that (after symmetrizing $a_i$)
\begin{align}\label{3ptobservation}
  \operatorname{tr}_\lambda \prod_{i = 1}^3 o(J^{a_i})q^{L_0^\text{crit}} = \mathfrak{d}(\lambda) d^{a_1 a_2 a_3} \operatorname{ch}_\lambda^\text{crit}(q) \ ,
\end{align}
where $\mathfrak{d}$ is a function of weight $\lambda$, and has nontrivial a limit $\mathfrak{d}_\text{3d}(\lambda) \equiv \lim_{\beta \to +0}\beta^3\mathfrak{d}(\beta^{-1}\lambda)$ as an odd function of $\lambda$.

Continuing with the same logic, one can apply the general Zhu's recursion formula to derive uniform differential operators that extract higher-point current correlation functions, and we conjecture the general wave function to be given by
\begin{align}
  \Psi_\lambda^{a_1 \ldots a_n}(z_i) = \langle J^{a_1}[z_1]\ldots J^{a_n}[z_n]\rangle^\text{crit}_\lambda \ .
\end{align}

In the case of $\mathfrak{g} = \mathfrak{su}(N)$, the 4d SCFTs admit surface defects engineered by the RG-flow initialized by some position dependent vev of baryonic Higgs branch operator \cite{Gaiotto:2012xa}.  As shown in \cite{Alday:2013kda}, these defects are labeled by the dominant integral weights $\lambda'$ of $\mathfrak{g}$, and the associated wave functions are given by $S_{\lambda \lambda'}/S_{\lambda 0}$. Here $S_{\lambda \lambda'}$ denotes the (analytically continued) modular S-matrix for $SU(N)$. Explicitly, $S_{\lambda \lambda'} = S_{00} \chi_\lambda(q^{\rho}) \chi_{\lambda'}(q^{\rho + \kappa})$ with $\rho = (\frac{N - 1}{2}, \ldots, - \frac{N - 1}{2})$, and $\kappa$ a set of integers linear defined by the Dynkin labels of $\lambda$. Hence, in the presence of $p$ such defects, the $n$-point correlation functions of currents $J^a$ are simply given by
\begin{align}
  & \ \operatorname{tr}_V  J^{a_1}[z_1] \ldots J^{a_n}[z_n]\prod_{\alpha = 1}^p \mathfrak{D}_{\lambda'_\alpha} q^{L_0}\\
  = & \ \sum_{\lambda} C_\lambda(q)^{2g - 2 - s} \Psi^{a_1 \ldots a_n}(z_i) \prod_{\alpha = 1}^p \frac{S_{\lambda \lambda'_\alpha}}{S_{\lambda 0}} \prod_{i = 1}^{s}\psi^{(i)}(q) \ .\nonumber
\end{align}


Nest we turn to a class of Argyres-Douglas theories denoted as $(J^{b}[\kappa], F)$ which admit class-$\mathcal{S}$ construction in terms of an irregular puncture and a full puncture \cite{Wang:2018gvb,Xie:2019yds,Song:2017oew}. Their associated VOAs are either given by, or contain as a vertex subalgebra, the affine algebra $V = \widehat{J}_{k = \frac{b}{b + \kappa} - h^\vee}$. For a class of such Argyres-Douglas theories, the Schur index can be computed from the 2d $q$-deformed Yang-Mills theory as \cite{Buican:2015ina,Cordova:2017mhb,Song:2017oew}
\begin{align}
  \operatorname{ch}_V(\mathbf{x}, q) = \sum_\lambda q^{- \frac{c}{24}} C_\lambda(q) q^{(\lambda, \rho)} \psi^{b, \kappa}_\lambda(\mathbf{a},q) \operatorname{ch}^\text{crit.}_\lambda(\mathbf{x}, q)\ .\nonumber
\end{align}
Here $\psi^{b, \kappa}_\lambda$ is the wave function associated to the irregular puncture with $\mathbf{a}$ the potential $U(1)^n$ flavor fugacities. The critical character is again associated to the full puncture with $\mathfrak{g}$ flavor fugacities $\mathbf{x}$.

Similar to the previous subsection, we can study the correlation functions of the currents employing Zhu's recursion formula. We begin by recalling  the recursion formula (\ref{twopointrecursion}) for two point functions. Note that the level $k$ is no longer critical. Treating the right hand side again as the differential operator $D^{ab}(z_1, z_2)$ acting on the index, which only acts on the fugacities of the full puncture but ignore the irregular puncture, we obtain the wave function
\begin{align}
  \Psi_\lambda^{ab}(z_1, z_2) & = \ K^{ab} \frac{|\lambda + \rho|^2 }{\dim \mathfrak{g}} \nonumber\\
  &  \ + K^{ab}\left[(k + h^\vee) E_2(\tau)+ \frac{ k \wp_0(z, \tau)}{(2\pi i)^2}\right] .
\end{align}
Note that the operator $D^{ab}(z_1, z_2)$ actually depends on the level $k$, and the $k + h^\vee$ term is the consequence of the mismatch between the critical level of the full puncture and the actual level $k$ of the full VOA.

To obtain the wave function of three point function, we simply apply the uniform differential operator $D^{a_1a_2}{_b}(z_1, z_2, z_3)$ on the two point wave function $\Psi^{ba_3}(z_2, z_3)$. Note that $D^{a_1a_2}{_b}(z_1, z_2, z_3)$ is independent of level $k$, and therefore the three-point wave function is simply
\begin{align}
  \Psi^{a_1 a_2 a_3}_\lambda(z_i)& \  = \langle J^{a_1}[z_1]\ldots J^{a_3}[z_3] \rangle^\text{crit}_\lambda + d^{a_1a _2 a_3} \text{ term} \nonumber\\ 
  & \ + \frac{b}{b + \kappa}D^{a_1a_2a_3}(z_1, z_2, z_3)\left(\frac{P_2(z_2 - z_3)}{(2\pi i)^2}\right) \ .\nonumber
\end{align}

Before moving on, let us comment on stress tensor insertions. In general the stress tensor of a theory of class $\mathcal{S}$ is not associated with any single puncture. Therefore there seems to be no notion of an associated wave function. This also manifests in the fact that the summand $\psi_\lambda$ of a general Schur index as a function of weight $\lambda$ is not an eigenfunction of $q \partial q$ which inerts $T$. However, it is easy to show that for $(J^{h^\vee}[\kappa], F)$ the full summand $\psi_\lambda$ with a given $\lambda$ is in fact an eigenfunction of $q \partial _q$,
\begin{align}
	\left(q \frac{d}{dq} - \frac{\dim \mathfrak{g}}{2}E_2(\tau) - \frac{|\lambda + \rho|^2}{2(k + h^\vee)}\right)\psi_\lambda = 0\ .
\end{align}
Therefore one may loosely associate the eigenvalue $\Psi_T \equiv \frac{\dim \mathfrak{g}}{2}E_2(\tau) - \frac{|\lambda + \rho|^2}{2(k + h^\vee)}$ with the ``wave function'' for $T$ insertions in these theories.

\section{Coulomb branch correlation functions in star-shape quiver}

In any three dimensions $\mathcal{N} = 4$ SCFT $\mathcal{T}^\text{3d}$, one can extract two protected subsectors forming two associative topological algebras, one in the Higgs branch and the other in the Coulomb branch. These algebras have been shown to be the deformation quantization of hyperk\"{a}hler cones describing the two branches of vacua \cite{Beem:2016cbd}.

One can derive such an algebra associated to the Higgs branch by performing suitable ``circle reduction'' or ``high termperture limit'' of the VOA $\mathbb{V}[\mathcal{T}^\text{4d}]$ where $\mathcal{T}^\text{4d}$ reduces to $\mathcal{T}^\text{3d}$ via the said reduction \cite{Pan:2019shz,Dedushenko:2019mzv}. More precisely, one puts $\mathcal{T}^\text{4d}$ on a twisted product $S^3 \times_q S^1$, where $q = e^{2\pi i \beta}$ with $\beta$ parametrizing the size of the circle factor. The $S^3\times_q S^1$-partition function of such a theory then computes the Schur index \cite{Pan:2019bor,Dedushenko:2019yiw}, while the correlation functions of suitably defined ``Schur operators'' inserted along a torus $T^2_q \subset S^3 \times_q S^1$ compute the torus correlation functions of the corresponding vertex operators in the VOA $\mathbb{V}[\mathcal{T}^\text{4d}]$.

The proper circle reduction sends $\beta \to 0$ while rescaling all Schur operators $\mathcal{O} \to \beta^h\mathcal{O}$ according to their conformal weights $h$. During such a reduction, only the 4d Schur operators belonging to the Higgs branch chiral ring survive, while others decouple in the sense that any correlation function involving these non-Higgs branch operators vanishes in the limit. The prime examples that survive are the moment maps operators of the 4d flavor symmetry.

The circle reduction $\mathcal{T}^\text{3d}$ of a $\mathcal{T}^\text{4d}$ is not necessary Lagrangian, while it always has a Lagrangian 3d mirror dual $\check{\mathcal{T}}^\text{3d}$. The Higgs branch correlation functions $\langle \mathcal{O}_\text{H}(x) \ldots\rangle$ of $\mathcal{T}^\text{3d}$ are mapped to those of $\check{\mathcal{T}}^\text{3d}$ in the Coulomb branch by the 3d mirror symmetry. It is therefore natural to expect the high-temperature limit of the current correlation functions of $\mathcal{T}^\text{4d}$ to reproduce a set of Coulomb branch correlators $\langle \mathcal{O}_\text{C}(x) \ldots\rangle$ in $\check{\mathcal{T}}^\text{3d}$. In the following we shall compute such topological Coulomb branch correlators for the 3d mirror dual of the $T_N$ theories, using the wave functions in the $q$-deformed Yang-Mills we have previously constructed.

We begin with the $T_2$ theory whose Schur index is given by
\begin{align}
  \mathcal{I}(x, q) = q^{- \frac{c}{24}} \sum_\lambda  C_\lambda(q)^2 \prod_{i = 1}^{3} q^{(\lambda, \rho)}\operatorname{ch}^\text{crit}_\lambda(\mathbf{x}^{(i)}, q) \ .
\end{align}
The naive limit of sending $\beta \to 0$ with fugacity identification $x = e^{2\pi i \xi}$ doesn't lead to a sensible outcome. Instead, following \cite{Nishioka:2011dq}, one should also rescale the Dynkin labels $\lambda_i \to \beta^{-1} \lambda_i$, originally denoting integral weights of the underlying algebra $\mathfrak{g}$, and make it continuous by replacing the sum $\sum_\lambda$ with an integral $\int \prod_{i = 1}^{r} d\lambda_i$. For the case at hand, $ r = 1$, and each critical character reduces to (up to overall factors involving only on $\beta$)
\begin{align}
  q^{(\lambda, \rho)}\operatorname{ch}_{\lambda}^\text{crit}(e^{2\pi i \xi}, e^{2\pi i \beta}) \xrightarrow{\beta \to 0} \frac{\sin 4\pi \lambda \xi}{\sinh(2\pi \lambda) \sinh(2\pi \xi)} \ , \nonumber
\end{align}
while
\begin{align}
  C_{\beta^{-1}\lambda}(e^{2\pi i \beta})^2 \to \frac{1}{2}\sinh^2(2\pi \lambda) = \Delta_{SU(2)}(\lambda) \ .\nonumber
\end{align}
reduces to the $SU(2)$ Haar measure factor. One immediately recognizes that the reduction of the critical character is nothing but the $S^3$ partition function of $T[SU(2)]$ theory with mass $\lambda$ and FI parameters $\xi$. The full index now becomes the matrix integral that computes the $S^3$-partition function of the star-shape quiver theory which fuses three $T[SU(2)]$ theories, with the $\lambda$ integral integrating over the VEV of the vector multiplet scalar which gauges the the manifest diagonal $SU(2)\subset SU(2)^3$ flavor symmetry of three $T[SU(2)]$ theories.

Now we extend the above computation to include rescaled current operators $\beta J^a[z]$ associated with one full puncture of $T_2$. First, we reduce the wave function $\beta^2\Psi_\lambda^{ab}(z, w)$, assuming $\operatorname{Re}(z - w) > 0$,
\begin{align}
  \beta^2\Psi_{\beta^{-1}\lambda}^{ab}(z, w) \xrightarrow{\beta \to 0} \frac{K^{ab}}{6}\left(1 + 4\lambda_1^2\right) \ .
\end{align}
We recognize this result to be the two-point function of a pair of magnetic monopoles, or two vector multiplet scalars in the massive $T[SU(2)]$ (identifying $m_1 = - m_2 = \lambda_1 $ while turning off the FI parameters) \cite{Dedushenko:2017avn},
\begin{align}
  \langle \mathcal{M}_+ \mathcal{M}_- \rangle^{S^3} = & \ \frac{1}{Z^{S^3}_{T[SU(2)]}} \int d\sigma \prod_{I = 1}^{2}\frac{i(\sigma + m_I )- 1/2}{2\cosh\pi (\sigma + m_I)} \nonumber \\
  = & \ \frac{1}{6}(1 + 4\lambda_1^2)\ .
\end{align}
The three-point wave function also has a nontrivial limit independent of the insertion coordinate,
\begin{align}
  \beta^3\Psi^{abc}_{\beta^{-1}\lambda}(z_i) \xrightarrow{\beta \to 0} - \frac{i f^{abc}}{12}(1 + 4\lambda_1^2) \ ,
\end{align}
which we can identify with the three-point function in the $T[SU(2)]$,
\begin{align}
  \langle \Phi \mathcal{M}_+ \mathcal{M}_- \rangle^{S^3}
  = & \ \frac{1}{Z^{S^3}} \int d\sigma \sigma \prod_{I = 1}^{2}\frac{i(\sigma + m_I) - 1/2}{2\cosh\pi (\sigma + m_I)} \nonumber \\
  = & \ - \frac{i}{12}(1 + 4\lambda_1^2)\ .
\end{align}
Finally, we identify the reduction of the current correlation function with the full correlation functions of said Coulomb branch operators in the star-shape quiver
\begin{align}
  & \ \operatorname{tr}_V J^a[z] \ldots J^c[0]q^{L_0 - \frac{c}{24}} \nonumber\\
  \to & \ \int d\lambda_1 \Delta_{SU(2)} \langle \mathcal{O}^a \ldots \mathcal{O}^c \rangle_{T[SU(2)]}^{S^3} (Z^{S^3}_{T[SU(2)]})^3 \ .
\end{align}

The above result immediately generalizes to $T_N$ theories. Indeed, the critical character reduces to the $S^3$-partition function of $T[SU(N)]$, while the correlation functions of the $SU(N)$ currents are expected to reduce to topological correlation functions $\langle \mathcal{O}^a \mathcal{O}^b \ldots\rangle$ of a set of Coulomb branch operators $\mathcal{O}^a$ in the $T[SU(N)]$ theory (again with vanishing FI-parameter $\xi$ but nonzero masses $\lambda_i$). For example,
\begin{align}
  q^{(\lambda, \rho)}\operatorname{ch}_\lambda^\text{crit}(\mathbf{x},q) \to & \ Z^{S^3}_{T[SU(N)]}(\lambda, \xi)\ , \nonumber\\
  \langle J^a[z]J^b[w]\rangle^\text{crit}_\lambda \to & \ \langle \mathcal{O}^a \mathcal{O}^b\rangle = K^{ab}\left(\frac{h^\vee}{12} + \frac{ |\lambda|^2}{\dim \mathfrak{g}}\right), \nonumber \\
  \langle J^a[z]J^b[w]J^c[0]\rangle^\text{crit}_\lambda \to & \ \langle \mathcal{O}^a \mathcal{O}^b \mathcal{O}^c \rangle\\
  = - \frac{i}{2}f^{abc}\bigg(\frac{h^\vee}{12} & \  + \frac{ |\lambda|^2}{\dim \mathfrak{g}}\bigg) + \mathfrak{d}_\text{3d}(\lambda)d^{abc} \ . \nonumber
\end{align}
Here $|\lambda|^2$ is the length-squared of the weight $\lambda$ with continuous Dynkin labels, which are identified with the masses of the $T[SU(N)]$ theories. The odd funciton $\mathfrak{d}_{3d}(\lambda)$ captures the high-temperature limit of the $d^{abc}$ term, which ultimately vanishes when the integral corresponding to the central gauge node in the star-shape quiver is performed.

The above computation is naturally compatible with the inclusion of surface defects. As discussed above, such defect introduces factors of modular S-matrices into the $q$-deformed Yang-Mills correlators. In performing the high-temperature limit, we send $\lambda_i \to \beta^{-1}\lambda_i$, meaning that the $\kappa$'s are also rescaled $\kappa_A \to \beta^{-1}\kappa_A$ for $A = 1, \ldots, N$. As a result, one simply has
\begin{align}
	\frac{S_{\lambda \lambda'}}{S_{\lambda 0}} \to \chi_{\lambda'}(e^{-2\pi \kappa_1}, \ldots e^{-2\pi \kappa_N})  \ .
\end{align}
This factor captues a Wilson loop in the central $SU(N)$ gauge node of the star-shape quiver. Geometrically, the Wilson loop links but does not insersect with the great circle where the Coulomb branch operators are inserted. Alternatively, this Wilson loop can be viewed as inserting a line operator into one of the $T[SU(N)]$ by acting on the $Z^{S^3}_{T[SU(N)]}$ with a difference operator $\mathfrak{D}_{\lambda'}$ labeled by the representation $\lambda'$ \cite{Alday:2013kda}. In particular, the line operator can be inserted into the same $T[SU(N)]$ theory in which the Coulomb branch operators are defined, since insertion of the Coulomb branch operators can be realized computationally by differentiating with respect to the FI-parameter (which is sent to zero in the end), and such differentiation commutes with the difference operator $\mathfrak{D}_{\lambda'}$ \cite{Dedushenko:2017avn}.


\vspace{1em}
In this letter we proposed a new way of computing current correlation functions on the torus via the $q$-deformed 2d Yang-Mills theory, and they are further identified with Coulomb branch correlation functions in the 3d mirror dual. The discussions in this letter can be generalized to moment map operators of generic regular punctures by following carefully the correlation functions along the qDS reduction \cite{Beem:2014rza}. Furthermore, it should be straightforward to derive the flavored correlation functions by a refined version of recursion relations \cite{Gaberdiel:2009vs}. With these generalizations, one can further compute from four dimensions the correlators of a large class of Coulomb branch operators in the 3d mirror dual. On $S^4$, a similar localization computation adapted to Schur-like operators can be performed \cite{Pan:2017zie}, which prompts the question that whether vertex operators in the Liouville/Toda theory can be defined that compute correlation functions of some of those Schur-like operators. A positive result would help further bridge the AGT-duality \cite{Alday:2009aq} and the SCFT/VOA correspondence.

\vspace{20pt}
\begin{acknowledgments}

We thank Wolfger Peelaers for previous collaborations which led to the current work, and for many illuminating discussions and suggestions. We also thank Yongchao L\"{u} and Jaewon Song for helpful discussions and comments. Y.P. is supported by the National Natural Science Foundation of China (NSFC) under Grant No. 11905301, the Fundamental Research Funds for the Central Universities under Grant No. 74130-31610023, the 100 Talents Program of Sun Yat-sen University under Grant No. 74130-18841207.
\end{acknowledgments}

\bibliography{reference}

\end{document}